\documentclass[12pt]{iopart}

\expandafter\let\csname equation*\endcsname\relax
\expandafter\let\csname endequation*\endcsname\relax

\usepackage{graphicx,amssymb,amsmath,amsthm,amsfonts,epsfig}
\usepackage[colorlinks=true]{hyperref}
\usepackage[usenames]{color}
\usepackage{epstopdf}

\usepackage{bm}
\usepackage{dcolumn}
\usepackage[utf8]{inputenc}
\usepackage{latexsym}
\usepackage{rotating}
\usepackage{xcolor}
\usepackage{longtable}
\usepackage{enumerate}
\usepackage{mathtools}
\usepackage{url}
\newcommand{\ben}{\begin{enumerate}}
\newcommand{\een}{\end{enumerate}}

\makeatletter
\renewcommand\footnoterule{%
  \kern-3\p@
  \hrule\@width2.5cm
  \kern2.6\p@}
\makeatother

\usepackage{mathrsfs}

\begin{document}

\title{From micro to macro and back: probing near-horizon quantum structures with gravitational waves}
\author{
Andrea Maselli$^{1}$,
Paolo Pani$^{1}$,
Vitor Cardoso$^{2,3}$,
Tiziano Abdelsalhin$^{1}$,
Leonardo Gualtieri$^{1}$,
Valeria Ferrari$^{1}$
}
\address{
$^{1}$Dipartimento di Fisica, Sapienza Università di Roma \& Sezione INFN Roma1, P.A. Moro 5, 00185, Roma, Italy\\
$^{2}$CENTRA, Departamento de F\'{\i}sica, Instituto Superior T\'ecnico -- 
IST, Universidade de Lisboa -- UL, Avenida Rovisco Pais 1, 1049 Lisboa, Portugal\\
$^{3}$Theoretical Physics Department, CERN 1 Esplanade des Particules, Geneva 23, CH-1211, Switzerland}

\begin{abstract}
Supermassive binaries detectable by the planned space gravitational-wave interferometer LISA 
might allow us to distinguish black holes from ultracompact horizonless objects, even for certain 
models motivated by quantum-gravity considerations.
We show that a measurement of a very small tidal Love number with $\approx 10\%$ accuracy 
(as achievable by detecting ``golden binaries'') may also allow us to distinguish between different 
models of these exotic compact objects, even when taking into account an intrinsic uncertainty 
in the object radius putatively due to quantum mechanics.
We argue that there is no conceptual obstacle in performing these measurements, the main 
challenge remains the detectability of small tidal effects and an accurate waveform modelling.
\end{abstract}

\maketitle

\section{Introduction}

Gravitational-wave~(GW) measurements of the tidal deformability of neutron stars~(NSs)~\cite{Flanagan:2007ix,Hinderer:2007mb} --~through 
the so-called tidal Love numbers~(TLNs)~\cite{PoissonWill}~-- provide the most accurate tool 
so far to probe the microphysics of the NS interior well above the nuclear saturation 
density~\cite{TheLIGOScientific:2017qsa,Hinderer:2010ih,Harry:2018hke,Abbott:2018exr,De:2018uhw, Abbott:2018wiz}.
The TLNs encode the deformation properties of a compact object, and describe 
how its multipole moments change in response to the external tidal field. 
The dominant contribution among the TLNs is given by the apsidal constant 
$k_2$, which characterizes quadrupolar deformations.
Two NSs with similar mass and radius --~but described by equations of state~(EoS) with 
different stiffness~-- can have a TLN that differs by as much as 100$\%$~\cite{Hinderer:2010ih}. 
The macroscopic difference in the TLNs acts as a \emph{magnifying glass} to probe the 
fundamental interactions within the NS core, for example to understand if the latter is made of 
normal $npe\mu$ matter, or hyperons, pion condensates, quarks, strange matter, 
etc~\cite{Lattimer:2006xb}.

It has been realized that GW measurements of the TLNs can also be used to distinguish black 
holes (BHs) from other ultracompact objects~\cite{Cardoso:2017cfl,Maselli:2017cmm,Sennett:2017etc,Barack:2018yly}.
The TLNs of a BH are identically zero~\cite{Binnington:2009bb,Damour:2009vw,Fang:2005qq,Gurlebeck:2015xpa,Poisson:2014gka,
Pani:2015hfa,Pani:2015nua,Pani:2015hfa}, whereas those of exotic compact objects~(ECOs) 
are small but finite~\cite{Pani:2015tga,Uchikata:2016qku,Porto:2016zng,Cardoso:2017cfl}.
Therefore, measuring a nonvanishing TLN with measurements errors small enough to exclude 
the null case would provide a smoking gun for the existence of 
new species of ultracompact massive objects~\cite{Cardoso:2017cfl,Maselli:2017cmm,Barack:2018yly,Johnson-McDaniel:2018uvs}.

Certain models of ECOs (all belonging to the ClePhO category introduced in 
Refs.~\cite{Cardoso:2017cqb,Cardoso:2017njb}, see below) are 
characterized by a TLN that vanishes as 
the logarithm of the (proper) distance (see Eq.~\eqref{k2log} below)
in the BH limit (i.e., when their radius $r_0$ tends to the Schwarzschild radius $2M$ in 
the $G=c=1$ units adopted hereafter). Owing to this  logarithmic behavior\footnote{This 
logarithmic behavior appears in various models of ECOs and it
is related to the emergence of the ``scrambling'' time~\cite{Hayden:2007cs,Sekino:2008he}. 
Similar logarithmic behaviors have been reported for other observables, e.g. GW 
echoes~\cite{Cardoso:2016rao,Cardoso:2016oxy,Cardoso:2017cqb,Cardoso:2017njb} and for the 
corrections to the multipole moments of certain ECO models relative to a Kerr 
BH~\cite{Pani:2015tga,Uchikata:2016qku}.}, the TLNs of ultracompact 
objects are still large 
enough to be measurable in the future~\cite{Cardoso:2017cfl,Maselli:2017cmm}, even for 
those models of ECOs which are motivated by quantum-gravity 
scenarios~\cite{Mazur:2004fk,Mathur:2005zp,Mathur:2008nj,Barcelo:2015noa, 
Danielsson:2017riq,Berthiere:2017tms}, in 
which case one expects $r_0\approx 2M+{\ell}_P$ (in a coordinate-independent way to 
be specified below; here ${\ell}_P\approx 1.6\times 10^{-33}\,{\rm cm}$ 
is the Planck length). In particular, it was pointed out that for 
highly-spinning supermassive binaries detectable by the future space interferometer 
LISA~\cite{Audley:2017drz} the signal-to-noise ratio might be high enough to distinguish BHs 
from ECOs even if the latter display Planckian corrections at 
the horizon scale~\cite{Maselli:2017cmm}.

The next most natural question, that we explore here, is the following: \emph{assuming 
such ECOs exist, would a future detection be able 
to distinguish among different models, possibly allowing for model selection of different 
quantum-gravity scenarios?}

\section{ECO model selection through TLNs}

In order to investigate the above question, we consider exotic, nonspinning objects with surface $r_0$ very close to 
the Schwarzschild radius. We parametrize such objects with a quantity\footnote{For the class of ClePhOs 
considered in this 
work --~i.e., those objects which feature a ``clean'' photon-sphere~-- the radius' shift is smaller than a certain 
threshold, namely $\delta/(2M)\lesssim0.0165$~\cite{Cardoso:2017cqb,Cardoso:2017njb}.} $\delta$, such that 
$r_0=2M+\delta$. One could also adopt the \emph{proper distance} $\Delta$ 
between the radius of the object and the would-be horizon~\cite{Abedi:2016hgu},
\begin{equation}
\Delta=\int_{2M}^{r_0}\frac{dr}{\sqrt{1-2M/r}}\approx\sqrt{8M\delta}\,,\label{eq:defdelta}
\end{equation}
where the last step is valid to leading order when $r_0\approx 2M$. We shall use 
$\Delta\to0$ as a coordinate-independent limit to the BH case. As we shall discuss, owing to the logarithmic dependence 
of the TLNs the distinction between $\Delta$ and $\delta$ is negligible.

The TLNs of three toy models of ultracompact objects which can be arbitrarily close to 
the compactness of a BH were computed in Ref.~\cite{Cardoso:2017cfl} by solving 
linearized 
Einstein's equations coupled to exotic matter fields, and with suitable boundary or 
junction conditions. For these classes of ECOs, $k_2$ scales logarithmically with the 
radius' shift $\delta$, namely $k_2\sim 1/\vert \log (\delta/M) \vert$. In 
terms of the proper distance $\Delta$, the (electric, quadrupolar) TLNs of these models in 
the limit $\Delta\ll 2M$ read
\begin{equation}
 k_2 \sim \left(a+b\log\left(\frac{\Delta}{4M}\right)\right)^{-1}\,, \label{k2log}
\end{equation}
where $a=(10,\frac{5(23-\log64)}{16},\frac{35}{8})$, 
$b=(\frac{15}{2},\frac{45}{8},\frac{15}{4})$ for wormholes, gravastars, and perfectly 
reflective objects, respectively. 
We consider these models as ECO prototypes for which the TLNs are known 
analytically. Indeed, the above logarithmic scaling is actually a rather general property. Extending the analysis of 
Ref.~\cite{Cardoso:2017cfl}, it is easy to show that the logarithmic scaling holds for any ECO whose exterior is 
Schwarzschild, and when generic Robin-type boundary conditions, namely $A\Psi+B\frac{d\Psi}{dr_*}=C$, 
are applied to the Zerilli function $\Psi$ at the surface (here $r_*$ is the standard tortoise coordinate). 
In this case, in the $\Delta\to0$ limit one gets
\begin{equation}
 k_2 \sim  \frac{8 A-6 C}{15 A} \log^{-1}\left(\frac{\Delta}{4M}\right)\,,
\end{equation}
the only exception concerns the zero-measure case $A=\frac{3}{4} C$, for which $k\sim 
\Delta^2/\log(\Delta/(4M))$. However, no ECO models described by these fine-tuned boundary conditions are known.
Note that, since $\Delta\sim\sqrt{\delta}$ and $k_2$ depends logarithmically on it, the distinction between $\Delta$ 
and $\delta$ only accounts for a factor of $2$ in the TLN.

\begin{figure}
	\centering		
\includegraphics[width=.4\textwidth]{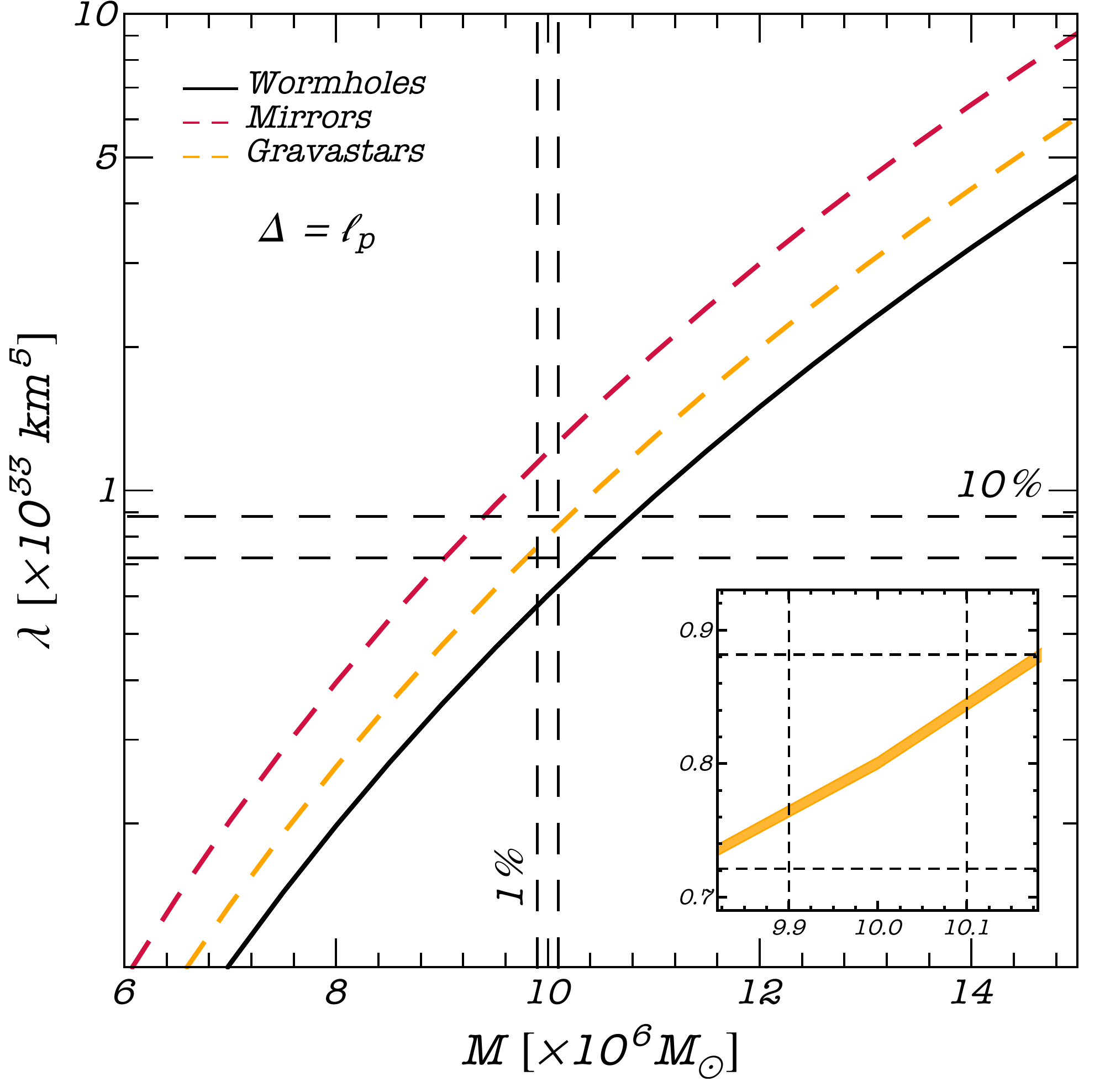}
	\caption{Tidal deformability $\lambda$ as a function of the mass for three 
different models of ECOs. For all models, the surface is at Planckian proper distance, 
$\Delta=\ell_P$, from the Schwarzschild radius. The dashed lines refer to a putative 
measurement 
of the TLN at the level of $10\%$ for an object with $M=10^7M_\odot$, which would allow 
to distinguish among different models. Although unnoticeable in 
the plot, each curve is actually a band of width $\ell_P$ to account for the intrinsic 
error due to the quantum uncertainty principle~\cite{Addazi:2018uhd} (the thickness is 
resolved only in the zoomed inset).}
	\label{fig:lambdaM}
\end{figure}

It has been recently argued that the exponential dependence of $\delta(k)$ and of 
its errors (see bands in Fig.~1 of Ref.~\cite{Maselli:2017cmm}) and the 
quantum uncertainty principle might prevent probing Planckian corrections at the 
horizon scale~\cite{Addazi:2018uhd}. We disagree with this conclusion.
Figure~\ref{fig:lambdaM} --~inspired by standard analysis to discriminate among 
NS equations of state~(EoS)~\cite{Hinderer:2010ih,Maselli:2013hl}~-- shows 
the tidal deformability $\lambda=\frac{2}{3}M^5|k_2|$ as a function of the 
object mass for the different models presented above. Crucially, in all cases we 
assume the emergence of a Planckian fundamental scale and set the proper 
distance $\Delta={\ell}_P$ (our results would be qualitatively the same if we consider $\delta=\ell_P$). 
This plots proves that the 
detectability of near-horizon quantum structures is not biased by any fundamental problem 
beside the observational challenge posed by extracting small TLNs from the GW signal. 
A putative measurement of $k_2\approx10^{-3}-10^{-2}$ with $10\%$ errors, as achievable for 
highly spinning LISA binaries up to luminosity distance 
of $2\,{\rm Gpc}$~\cite{Maselli:2017cmm}, would allow to distinguish among all three models at more than $90$\% 
confidence level.\footnote{We refer the reader to \cite{Maselli:2017cmm} for a detailed analysis on the statistical 
errors and on the systematics related to the TLN's measurements by GW interferometers.}
Thus, even though the microscopic scale of the correction, $\Delta={\ell}_P$, is the same 
for all models, the TLNs (i.e., the macroscopic quantities that 
really enter the waveform) are different enough to allow for discrimination.

In Fig.~\ref{fig:lambdaM} we have also included the intrinsic error 
coming from the quantum uncertainty principle as proposed in 
Ref.~\cite{Addazi:2018uhd}. This implies an intrinsic uncertainty on length scales at 
the level of $\ell_P$ for energies of the order of the Planck mass. Since the latter 
is enormously smaller than the mass $M$ of these objects, one would expect that the 
effect of the quantum uncertainty principle is negligible. This is confirmed by 
Fig.~\ref{fig:lambdaM}, where each curve is actually a very narrow
band obtained by considering $\delta={\ell}_P\pm {\ell}_P/2$, i.e. with an intrinsic uncertainty $\pm 
{\ell}_P/2$~\cite{Addazi:2018uhd}. 
This effect is negligible compared to the statistical errors on $\lambda$. 
This result is also consistent with Fig.~1 in Ref.~\cite{Addazi:2018uhd} and with the fact that 
the wavelengths relevant for this system are of the order of the orbital separation of 
the binary, $d\gtrsim {\cal O}(M)\sim {\cal O}(10^{45})\left(\frac{M}{10^7M_\odot}\right) 
\ell_p$ at least.

\section{Probing quantum structures at the horizon?}

Our results confirm and extend the analysis of Ref.~\cite{Maselli:2017cmm}, suggesting 
that not only should it be possible to use future GW measurements of the TLNs to 
distinguish between BHs and ECOs (even for those ECO models in which $\Delta={\ell}_P$), 
but also that --~with a slightly higher signal-to-noise ratio~-- it might be possible to 
distinguish between different ECO models all with 
$\Delta=\ell_P$ but with different TLNs. 
This might allow to perform ECO model selections 
and possibly rule out certain scenarios that predict a particular 
ECO rather than another.

This tantalizing possibility should not come as a surprise, since this is 
precisely the same strategy used to constrain the NS EoS from GW measurements of the 
TLNs~\cite{TheLIGOScientific:2017qsa,Hinderer:2010ih,Harry:2018hke,Abbott:2018exr,
De:2018uhw, Abbott:2018wiz}. One might argue that, since different EoS differ by the 
microscopic interactions occurring above the QCD scale --~roughly $200\,{\rm 
MeV}$ or $1\,{\rm fm}$~-- one would need a ``gravitational microscope'' with such length 
resolution~\cite{Addazi:2018uhd}. If correct, this line of reasoning would prevent any 
constraint on the NS EoS through the TLNs, since the resolution on the wavelength of the 
GW signal from compact binaries is not even microscopic. The key point is that 
\emph{microscopic} effects acting at small scales lead to different \emph{macroscopic} 
TLNs; the latter are the quantities effectively entering the waveform and therefore 
measurable.

Another example of the magnification of quantum effects in compact stars is provided by 
Chandrasekhar's mass limit~\cite{Chandrasekhar:1935zz},
\begin{equation}
 M_{\rm Ch}\sim \frac{M_P^3}{m_H^2}\,,
\end{equation}
where $M_P=\ell_P c^2/G$ and $m_H$ are the Planck mass and the mass of the proton, 
respectively. Since $\ell_P=\sqrt{\hbar G/c}$, a hypothetical change of the fundamental 
quantum scale governing the microphysics of the object would affect the Chandrasekhar 
mass \emph{macroscopically}. For the sake of the argument, if (say) $\hbar\to 2\hbar$, 
then $M_{\rm Ch}\propto \hbar^{3/2}$ would change roughly by a factor of $3$. Likewise, 
a putative intrinsic error $\ell_P\pm \ell_P/2$ would affect the Chandrasekhar limit at 
the level of kilometers.

Thus, at variance with Ref.~\cite{Addazi:2018uhd}, we argue that there is no 
fundamental or conceptual obstacle in probing Planckian corrections at the horizon scale 
\footnote{We remark that for ECOs which do not feature Planckian corrections (such as boson stars, which have a maximum 
compactness $M/r_0\sim 0.3$), the TLNs are much bigger and easier to measure. In such case LISA would 
be able to measure TLNs from GW observations at the level of 1\% and below~\cite{Cardoso:2017cfl}.}.

The real challenge is on the detectability side and 
parameter estimation, due to the smallness of the tidal deformability for these ECO 
models~\cite{Maselli:2017cmm}, the systematics of the waveform 
modeling~\cite{Favata:2013rwa,Wade:2014vqa,Samajdar:2018dcx,Harry:2018hke}, and the 
requirement to reach a \emph{resolution in the TLNs} small enough to distinguish two ECO 
models with $\Delta\approx \ell_P$.
Future detectors seem on the verge to be able to detect this effect, the final answer 
will depend on the uncertain event rates and on the ability of building accurate 
waveforms.\\

\begin{ack}
AM and PP acknowledge financial support provided under the European Union's H2020 ERC, Starting 
Grant agreement
no.~DarkGRA--757480.
V.C. acknowledges financial support provided under the European Union's H2020 ERC 
Consolidator Grant ``Matter and strong-field gravity: New frontiers in Einstein's theory'' 
grant agreement no. MaGRaTh--646597. 
This project has received funding from the European Union's Horizon 2020 research and 
innovation programme under the Marie Sklodowska-Curie grant agreement No 690904.
We acknowledge financial support provided by FCT/Portugal through grant PTDC/MAT-APL/30043/2017.
We acknowledge the SDSC Comet and TACC Stampede2 clusters through NSF-XSEDE Award Nos. PHY-090003.
The authors would like to acknowledge networking support by the GWverse COST Action 
CA16104, ``Black holes, gravitational waves and fundamental physics.''
We acknowledge support from the Amaldi Research Center funded by the 
MIUR program ``Dipartimento di Eccellenza''~(CUP:~B81I18001170001).
\end{ack}

\bibliographystyle{iopart-num}
\bibliography{refs}

\end{document}